\documentclass[12pt]{article}
\input epsf

\newcommand{\be}{\begin{equation}}
\newcommand{\ee}{\end{equation}}

\newcommand{\beq}{\begin{equation}}
\newcommand{\eeq}{\end{equation}}
\newcommand{\bea}{\begin{eqnarray}}
\newcommand{\eea}{\end{eqnarray}}
\newcommand{\beal}{\setcounter{letter}{1} \begin{eqnarray}}
\newcommand{\eeal}{\addtocounter{equation}{1} \end{eqnarray}}

\newcommand{\req}[1]{Eq.(\ref{#1})}

\begin{document}


\title{\bf The 5-D Choptuik critical exponent and holography}

\author{J. Bland${}^1$ and G. Kunstatter${}^2$}

\maketitle

{\it 
\begin{center}${}^1$ Dept. of Physics and Astronomy\\
and Winnipeg Institute of
Theoretical Physics\\
University of Manitoba, Winnipeg, Manitoba Canada R3T 2N2.\\[5pt]
\end{center}
\begin{center}${}^2$ Dept. of Physics and Winnipeg Institute of
Theoretical Physics\\
University of Winnipeg, Winnipeg, Manitoba Canada R3B 2E9.\\[20pt]
\end{center}
}

\begin{abstract}
Recently, a holographic argument was used to
relate the saturation exponent, $\gamma_{BFKL}$, of four-dimensional
Yang-Mills theory in the Regge limit to the Choptuik critical
scaling exponent, $\gamma_{5d}$, in 5-dimensional black hole
formation via scalar field collapse~\cite{alvarez-gaume}. Remarkably, the 
numerical value of the former agreed quite well with previous
calculations of the latter. We present new results of an improved
calculation of $\gamma_{5d}$ with substantially decreased 
numerical error. Our current result is $\gamma_{5d} = 0.4131 \pm
0.0001$, which is close to, but not in strict agreement with, the
value of $\gamma_{BFKL}=0.409552$ quoted in \cite{alvarez-gaume}.
\end{abstract}


The critical behaviour exhibited by gravitational collapse to black
holes, first discovered by Choptuik~\cite{choptuik} (see
\cite{gund,lehn} for reviews), is a small scale, strong
gravitational field effect that is by now well understood at the
qualitative level. As well, key physical quantities such as the
associated critical solution and scaling exponents have been
calculated numerically. Precise analytic derivations of these
quantities are still, to the best of our knowledge, lacking.

Recently, intriguing holographic arguments were used to relate the
strong field gravitational scaling behaviour in spherically symmetric
scalar field collapse in five space-time
dimensions to scaling in a suitable, 4-D weakly coupled gauge
theory~\cite{alvarez-gaume}. Specifically, it was conjectured that
exponential growth of the scattering amplitude for two hadrons in
the one-pomeron exchange approximation corresponded via a
holographic mapping to the exponential growth away from the critical
solution that leads to Choptuik scaling in five dimensions. Thus, it
was argued, the BFKL scaling exponent $\gamma_{BFKL}$ should equal
the 5-dimensional Choptuik scaling exponent $\gamma_{5d}$.

As shown by the authors of~\cite{alvarez-gaume}, a numerical
calculation of the BFKL scaling exponent yields the value
$\gamma_{BFKL}=0.409552$. On the other hand, the most accurate 
previous calculation yields a 5-D
Choptuik scaling exponent of: 
$\gamma_{5d} = 0.412 \pm 0.004$ \cite{bland1}. A somewhat lower result
, namely $0.408$, was reported  
in~\cite{sorkin}, but with an error of $\pm 0.008$. 
The apparent agreement between $\gamma_{BFKL}$ and
$\gamma_{5d}$ provides startling support for the
conjecture relating 5d gravity to pomeron exchange in 4d Yang-Mills
theory, and for the general validity of the holography hypothesis.

The purpose of this note is to present the results of a
more accurate calculation of $\gamma_{5d}$ and the related echoing
period $\Delta_{5d}$. 
Recall that the Choptuik scaling exponent is most directly 
observed in the simple scaling law obeyed by the horizon radius near
criticality~\cite{choptuik} \beq
R_{AH}\propto(a-a_*)^{\gamma}. 
\label{scaling}
\eeq 
where $R_{AH}$ is the horizon radius on formation, $a$ is a parameter
describing the initial data whose critical value $a_*$ separates initial data that
lead to blackhole formation from those that lead to dispersal of the scalar field. The echoing period $\Delta_{5d}$ derives from the discrete self-similarity of the critical solution, and can
be extracted from the periodicity of the critical scalar field solution:
\beq
\chi(r,t) = \chi\left(re^{\Delta_{5d}},te^{\Delta_{5d}}\right)
\eeq
The new values we obtain for these quantities are 
\beq \gamma_{5d} =
0.4131\pm 0.0001 \hspace*{25pt} \Delta_{5d} = 3.225 \pm 0.003.
\label{gamma} \eeq 
These results are consistent within error with previous
calculations, but $\gamma_{5d}$ no longer agrees precisely with
$\gamma_{BFKL}$. Nevertheless, given the uncertainties in the
calculation of $\gamma_{BFKL}$, a discrepancy of this order (less
than 1$\%$) is perhaps not surprising\cite{a-g_private}.

We now describe how (\ref{gamma}) was
obtained. Our formalism and code are basically the same as that used
in \cite{bland1}, which in turn was based on the analysis of
\cite{d_dim1}. A couple of crucial changes have permitted the
increase in the accuracy by a factor of about twenty.

The calculation can generically be formulated in terms of Einstein
gravity in $D$ space-time dimensions minimally coupled to a massless
scalar field\cite{d_dim1}: 

\bea S^{(D)}&=&{1\over 16\pi
G^{(D)}}\int d^Dx\sqrt{-g^{(D)}}R(g^{(D)}) \nonumber \\
   && - \int d^Dx\ \sqrt{-g^{(D)}}|\partial\chi|^2.
\label{Einstein}
\eea
Spherical symmetry is imposed by requiring: 
\bea ds^2_{(D)} &=&
\bar{g}_{\alpha\beta} dx^\alpha dx^{\beta} + r^2(x^\alpha)
d\Omega_{(D-2)},\nonumber\\
\chi&=& \chi(x^\alpha)
\label{metric 1}
\eea
where $d\Omega_{(D-2)}$ is the metric on $S^{D-2}$ and $\alpha,\beta =
1,2$.
A relatively simple set of equations can be obtained by the following field
redefinition, which is motivated by 2-dimensional dilaton
gravity~\cite{d_dim1,dil_grav}: 
\bea
\phi &:=& {D-2\over 8(D-3)}\left({r\over l}\right)^{D-2}, \\
g_{\alpha\beta} &:=& \phi^{(D-3)/(D-2)}\ \bar{g}_{\alpha\beta},
\label{defs} \eea 
where $r$ is the optical scalar and $\phi$ is
proportional to the area of a $(D-2)$-sphere at fixed radius $r$.

We go to double null coordinates first introduced by
Garfinkle~\cite{DG}: 
\beq ds^2 = - 2 l g(u,v)\phi'(u,v) du dv
\label{double null} \eeq in which the relevant field equations take
the form: \bea & &\dot{\phi}' = - {D-3\over 2} \left({r\over
l}\right)^{D-3\over D-2} g \phi'
\label{double null equations a}\\
& &{g'\phi'\over g \phi} = 32\pi (\chi')^2 \label{cons}\\
& &(\phi\chi')^{\cdot} + (\phi\dot{\chi})' = 0, \label{double null
equations c} \eea 
where the prime and dot denote $v$ and $u$
derivatives, respectively. We henceforth restrict attention to five
dimensions by setting $D=5$, in which case $r \propto \phi^{1/3}$.

\begin{figure}[hbtp]
\begin{center}
\epsffile{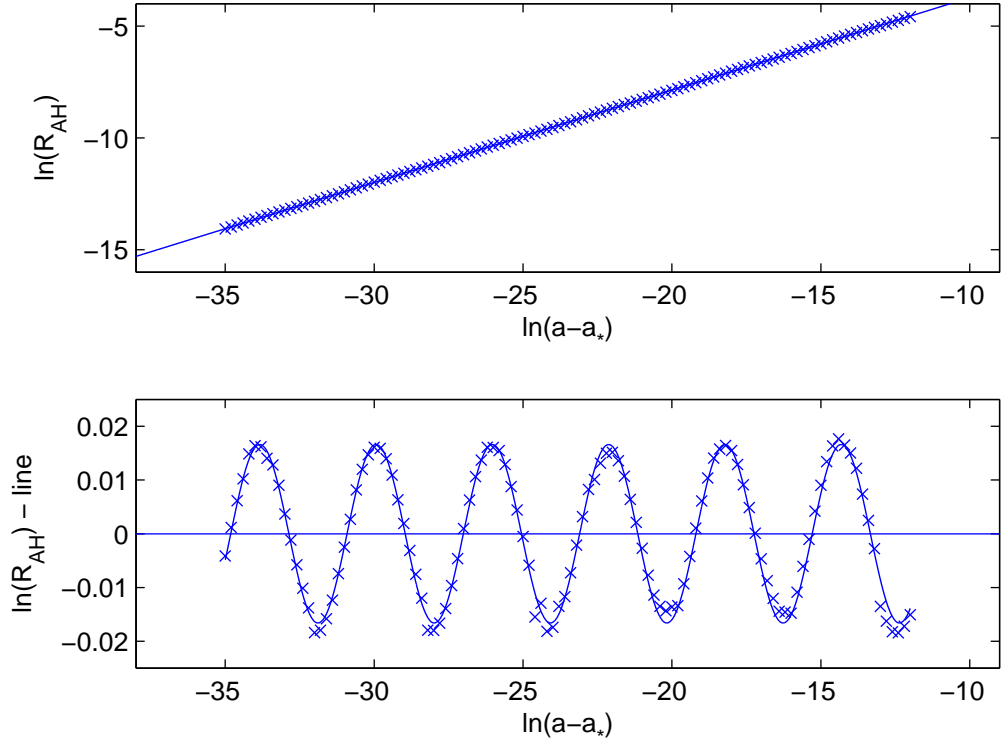} \caption{Plot of $ln(R_{AH})$ vs
$ln(a-a_*)$ (top graph). The value of $a_*$ was adjusted within a
small range in order to increase the quality of the fit and to
determine the error in $\gamma_{5d}$. In the bottom graph, the
linear portion of the fit is subtracted from $ln(R_{AH})$. The
residuals are well described by a sine wave (solid curve) and
clearly show the oscillation expected due to the discrete
self-similarity of the critical solution.} \label{5d_gamma}
\end{center}
\end{figure}

\begin{figure}[hbtp]
\begin{center}
\epsffile{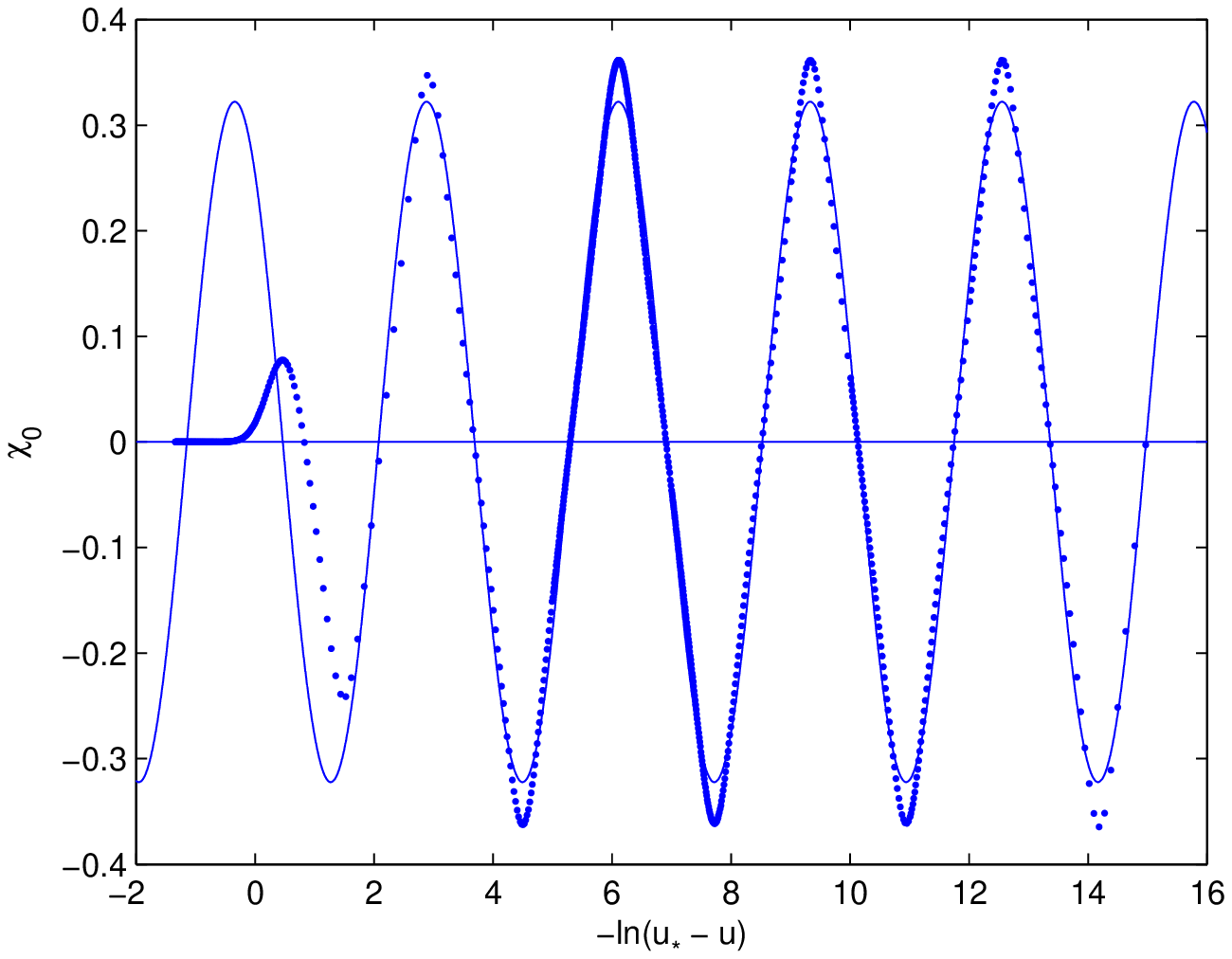} \caption{Plot of the matter field at the
origin vs $-ln(u_* - u)$. The value of $u_*$ was adjusted within a
small range to increase the quality of the fit and to determine the
uncertainty of the best estimate of the self-similarity period.
$\Delta_{5d}$ is determined by calculating the period of the
oscillation (the solid line represents the best fit to the data.)}
\label{5d_Delta}
\end{center}
\end{figure}

The general method we use 
starts with the choice of a one parameter family of initial data for
the scalar field $\chi$. Coordinate invariance allows the initial
value of $\phi$ to be chosen arbitrarily, while the  metric
component $g$ is obtained by integrating the constraint \req{cons}.
We then evolve the data using the remaining dynamical equations
(\ref{double null equations a}) and (\ref{double null equations c})
until either a horizon forms, or the field disperses. Horizon
formation is signalled by the vanishing of the null expansion
scalar, which for spherically symmetric geometries is given by: 
\be AH \equiv
{g}^{\alpha\beta}\partial_\alpha
 \phi \partial_\beta \phi.
\label{aheqn} \ee 
This calculation is repeated for different
values of some parameter, $a$, describing the initial data in order
to determine the critical value, $a_*$, separating black hole final
states from dispersed final states. Once $a_*$ is determined, we
calculate the horizon radius at formation, $R_{AH}$, for a wide
range of super-critical values of $a$ near $a_*$ and then plot 
$\ln(R_{AH})$ vs $\ln(a-a_*)$. This
generically yields an
(almost) straight line (see Figure \ref{5d_gamma}) from which the scaling
exponent and echoing period can be obtained by fitting 
the ln-ln plot to the following 5 parameter
function: 
\be \ln \left( R_{AH} \right) = c_0 +
\gamma_{5d} \ln(a-a_*) + c_1 \sin \left[ \left( {2 \pi \over
\delta_{5d}} \right) \ln(a-a_*) + c_2 \right], \label{eq:fit} \ee
where $\{c_0,c_1,c_2\}$ are family dependent constants. The
periodic deviations from a line in the ln-ln plot yield the
echoing period via the relation $\Delta_{5d} = 2 \gamma_{5d}
\delta_{5d}$~\cite{hod97}. A more accurate value for $\Delta$
can be obtained by measuring the period of oscillations of the
matter field at the origin as a function of $-ln(u_*-u)$, where
$u_*$ is the $a = a_*$ value of the coordinate $u$ at collapse (See
Figure \ref{5d_Delta}).

Our specific numerical scheme used a $v$ (`space') discretization to obtain a
set of coupled ODEs: \be h(u,v) \rightarrow h_i(u) \hspace*{25pt}
\phi(u,v)\rightarrow \phi_i(u) \ee where $i = 0,\ldots,(N-1)$
specifies the $v$ grid and $h = \chi + 2 \phi
\chi^{\prime}/\phi^{\prime}$. Initial data for these two functions
were prescribed on a constant $v$ slice, from which the function
$g(u,v)$ was constructed.

In all cases we started with Gaussian initial data 
\be 
\chi(u=0,\phi) = a
\phi^{2/3}\ {\rm exp}\left[-\left(\ {\phi-\phi_0\over
\sigma}\right)^2\right]. 
\ee 
The initial scalar field
configuration $\chi(u=0,\phi)$ is most conveniently specified as a
function of $\phi$ rather than $r$. (Recall that $\phi \propto
r^3$.) The value of the dilaton field at each grid
point on the initial surface was chosen for simplicity to be: 
\be
\phi_i(u=0,v) = (i+1) \Delta v, \ee 
where $\Delta v$ is the initial
grid spacing. 
The initial values of the other
functions were determined in terms of the above by computing the
integrals $g_i$  using Euler's method for
equally spaced points.

The results in \req{gamma} were obtained with an initial spacing $\Delta v
= 0.0012$ along with an initial grid size of $N = 2048$. We further
verified these results by performing calculations at several
smaller grid sizes (with proportionately larger initial grid
spacings) (See Table \ref{tab:summary}). To the stated error, the scaling exponent was independent of initial grid size, while echoing period $\Delta_{5d}$
 varied between about 3.223 and 3.226. This  determined the error bars on $\Delta_{5d}$ given in \req{gamma}.
\begin{table}[hbtp]
\begin{center}
\begin{tabular}{||c||c||c||c||} \hline
 Initial Grid Size & Initial Grid Spacing & $\gamma_{5d}$ & $\Delta_{5d}$  \\ \hline \hline
 $512$  & $0.0048$   & $0.4130 \pm 0.0001$ & $3.2264 \pm 0.0005$ \\ \hline
 $800$  & $0.003072$ & $0.4130 \pm 0.0001$ & $3.2230 \pm 0.0006$ \\ \hline
 $1024$ & $0.0024$   & $0.4131 \pm 0.0001$ & $3.2223 \pm 0.0006$ \\ \hline
 $1536$ & $0.0016$   & $0.4131 \pm 0.0001$ & $3.2264 \pm 0.0004$ \\ \hline
 $2048$ & $0.0012$   & $0.4131 \pm 0.0001$ & $3.2246 \pm 0.0005$ \\ \hline \hline
\end{tabular}
\end{center}
\caption{Choptuik 5-d critical exponent and echoing period calculated
with varying initial grid sizes.} 
\label{tab:summary}
\end{table}

Evolution in the `time' variable $u$ was performed using the $4^{th}$
order Runge-Kutta method. The general scheme is similar to that used
in~\cite{GP}, together with some refinements used in~\cite{DG}. This
procedure was also used for the $3-$dimensional collapse
calculations in~\cite{husain}. The accuracy of the calculation was
improved by decreasing the time steps as the calculation progressed
according to the formula: 
\be
\Delta u = \Delta u_0 \left( 1 - 0.999 \exp \left[ - \left( {u_* - u
\over 0.1} \right)^2 \right] \right), \ee 
where $\Delta u_0 = 0.001$
is the initial time step spacing and $u_*$ is the estimated value of
the coordinate $u$ of the critical solution. Further improvements in
accuracy were made by decreasing the value of $\Delta u_0$
periodically during the calculation, or once apparent horizon
formation commenced, i.e., once the function $AH(\phi)$ developed an
extremum. Time steps ranged from $\Delta u = 10^{-3}$ at the
beginning of the calculation to about $\Delta u = 10^{-11}$ near the
end.

The boundary conditions at fixed $u$ are 
\be \phi_k=0  \hspace*{15 pt} g_k=1,
 \ee 
where $k$ is the index
corresponding to the position of the origin $\phi=0$. All grid
points $ 0\le i \le k-1$ correspond to ingoing rays that have
reached the origin and are dropped from the grid. The boundary
conditions are equivalent to $r(u,u)=0$, $g|_{r=0} = g(u,u) =1$.
Notice that for our initial data, $\phi_k$ and hence $h_k$, are
initially zero, and therefore remain zero at the origin because of
Eq.(\ref{double null equations c}). 

The most significant improvement in the numerical code over~\cite{bland1}, involved systematically decreasing the grid spacing
$\Delta v$ using a method similar to that of \cite{DG}. 
At regular intervals in $ln(u_* - u)$, the exterior half of the gridpoints were discarded. The number of gridpoints was then restored
 by using a $4^{th}$ order Lagrange interpolation between the remaining grid
points, thereby doubling the resolution.  Typically, by the end of the calculation
the grid spacing was decreased to approximately $10^{-6}$. This allowed the accuracy of the horizon position to be maintained
over the entire ln-ln plot and reduced significantly the
overall uncertainty in $\gamma_{5d}$.

At each $u$ step, a check was made to see if an apparent horizon had
formed by observing the function $AH$ in Eq.(\ref{aheqn}) whose
vanishing signals the formation of an apparent horizon. For each run
of the code with fixed parameters $a$, $\sigma$, and $\phi_0$, this
function was scanned from larger to smaller radial values after each
Runge-Kutta iteration for the presence of an apparent horizon.  In
the sub-critical case, all the radial grid
points reached zero without detection of an apparent horizon, which
signals  pulse reflection. In the super-critical case, black
hole formation was signalled by the vanishing of $AH$ at some finite radius.
Since the code crashes when $AH<0$, so the 
the time evolution was slowed down as $AH \rightarrow 0^+$ in order to obtain the most
accurate estimate of the apparent horizon position. For any given run, 
the apparent horizon position ($R_{MIN}$) was
assigned the value of the grid point for which $AH$ reached a
minimum during the last surviving iteration. 

For our initial runs we fixed the location and width of the
Gaussian initial data to be $\phi_0=1$ and $\sigma =0.3$, respectively.
We then varied the amplitude $a$ to determine the
critical amplitude $a_*$ via a binary search. The code was written
using double precision in C thus, the binary search was terminated
when $a_*$ was determined to $15$ digits. We then did runs in the
super-critical region $a>a_*$ to determine the dependence of the
horizon radius $R_{AH}$ on the amplitude. It was found that
numerical error would introduce noise in the ln-ln plot for the runs
closest to $a_*$ and so only super-critical runs for which $(a -
a_*)/a_* > 10^{-13}$ were fitted to determine $\gamma_{5d}$. To
improve the quality of the fit Eq. (\ref{eq:fit}), the value of
$a_*$ was adjusted within the range of values between super-critical
and sub-critical collapse obtained during the binary search. The
error associated with $\gamma_{5d}$ represents the range of fit
results when $a_*$ is adjusted within that range. The universality
of the critical solution was verified by starting runs $a=a_*$ and varying
the width and center of the initial data profile. As shown in Table
\ref{tab:summary2}, the results for $\gamma_{5d}$ agreed within error
in all cases.
\begin{table}[hbtp]
\begin{center}
\begin{tabular}{||c||c||} \hline
 Critical parameter & $\gamma_{5d}$  \\ \hline \hline
 $a$      & $0.4131 \pm 0.0001$ \\ \hline
 $\sigma$ & $0.4131 \pm 0.0003$ \\ \hline
 $\phi_0$ & $0.4133 \pm 0.0004$ \\ \hline \hline
\end{tabular}
\end{center}
\caption{Choptuik scaling exponent in 5 dimensional super-critical
collapse. We verified the universality of the critical solution by
varying the width and center of the initial data profile. The
scaling constants calculated by varying $\sigma$ and $\phi_0$ are in
agreement with the results obtained by varying the amplitude of the
initial pulse. Universality was tested at an initial grid size of
$2048$.} \label{tab:summary2}
\end{table}

To summarize, we have obtained fairly accurate values for the
Choptuik scaling exponent and echoing period in spherically
symmetric scalar field collapse in five dimensions: 
\beq \gamma_{5d}
= 0.4131\pm 0.0001 \hspace*{25pt} \Delta_{5d} = 3.225 \pm 0.003
\label{gamma2}. 
\eeq
These results agree within error with all previous calculations, but
the value of $\gamma_{5d}$ does differ somewhat from
$\gamma_{BFKL}$. The discrepancy is small but potentially
significant. It should, however, be kept in mind that the value
calculated in~\cite{alvarez-gaume} is strictly speaking exact only
in the asymptotic high energy region and it is not yet completely
clear where this region begins. Hence there may be corrections
due to the usual running of the strong coupling constant, saturation
and other effects~\cite{a-g_private}. Thus although the proposed holographic connection
between $\gamma_{BFKL}$ and $\gamma_{5d}$ is compelling, more work is required
in order to confirm its validity.\\[10pt]

\noindent
{\bf Acknowledgements}\\
This work was supported in part by the Natural Sciences and Engineering
Research Council of Canada. We are thankful to L. Alverez-Gaume and D. Garfinkle
for useful discussions and encouragement.

\par

\end{document}